\documentclass[twocolumn,superscriptaddress,amsmath,aps]{revtex4}
\usepackage{graphicx,color}
\usepackage{bm}
\usepackage[hypertex]{hyperref}
\usepackage{CJK}
\usepackage{booktabs} 
\usepackage{array}
\usepackage{float}

\newcommand{\be}{\begin{equation}}
\newcommand{\ee}{\end{equation}}
\newcommand{\bea}{\begin{eqnarray}}
\newcommand{\eea}{\end{eqnarray}}
\newcommand{\bsube}{\begin{subequations}}
\newcommand{\esube}{\end{subequations}}

\newcommand{\Eq}[1]{Eq.\,(\ref{#1})}
\newcommand{\Eqs}[1]{Eqs.\,(\ref{#1})}

\newcommand{\la}{\langle}
\newcommand{\ra}{\rangle}

\newcommand{\nl}{\nonumber \\}

\usepackage{amsfonts}
\usepackage{amsmath}
\usepackage{amssymb}
\usepackage{graphicx}
\usepackage{type1cm}
\usepackage{times}
\usepackage{bm}
\usepackage{color}

\newcommand{\beq}{\begin{equation}}
\newcommand{\eeq}{\end{equation}}
\newcommand{\beqn}{\begin{eqnarray}}
\newcommand{\eeqn}{\end{eqnarray}}
\newcommand{\bsub}{\begin{subequations}}
\newcommand{\esub}{\end{subequations}}

\begin{document}
\begin{CJK}{GBK}{song}


\title{Optical phase estimation via homodyne measurement
in the presence of saturation effect \\  of photodetectors }

\author{Jialin Li}
\affiliation{Center for Joint Quantum Studies and Department of Physics,
School of Science, \\ Tianjin University, Tianjin 300072, China}
\author{Yazhi Niu}
\affiliation{Center for Joint Quantum Studies and Department of Physics,
School of Science, \\ Tianjin University, Tianjin 300072, China}

\author{Lupei Qin }
\email{qinlupei@tju.edu.cn}
\affiliation{Center for Joint Quantum Studies and Department of Physics,
School of Science, \\ Tianjin University, Tianjin 300072, China}

\author{Xin-Qi Li}
\email{xinqi.li@imu.edu.cn}
\affiliation{Research Center for Quantum Physics and Technologies,
Inner Mongolia University, Hohhot 010021, China}
\affiliation{School of Physical Science and Technology,
Inner Mongolia University, Hohhot 010021, China}


\date{\today}

\begin{abstract}
{\flushleft For optical}
phase estimation via homodyne measurement, we generalize the theory 
from detector's linear to nonlinear response regime,
which accounts for the presence of saturation effect.
For optical coherent light, we carry out analytic expressions 
for detector's current and estimate precision.
Using specific device parameters, we illustrate
the improved estimation after accounting for the saturation effect.
\end{abstract}


\maketitle

{\flushleft Optical phase estimation}
is at the center of many fundamental and practical problems,
where the information of parameter is encoded in the phase of
an electromagnetic field \cite{Hel76,Sch93,WM09}.
In optical phase measurement, the precision is usually bounded by the standard quantum
limit (SQL), say, with a precision scaling as $1/\sqrt{N}$ \cite{Llo04},
where $N$ is the average photon number of the optical field.
In order to enhance the metrology precision, the most direct way
is increasing the intensity of light, such as in LIGO.
Some recent studies considered to improve further the sensitivity
by proposing a nonlinear Mach-Zehnder interferometer
--introducing Kerr nonlinearities into the laser interferometer--
for detecting relativistic gravity effects, 
such as estimating the Schwarzschild radius of a black hole
or quantum detection of possible wormholes \cite{Ra17,Sab17,Sab18}.
In the large $N$ limit, the precision can scale with
the photon numbers by $\sim 1/N^{1.5}$,
which goes beyond the Heisenberg limit ($\sim 1/N$).
For the purpose of obtaining the Kerr nonlinearity advantage,
$10^{15}\sim 10^{20}$ photons per optical pulse are required \cite{Ra17,Sab17,Sab18}.
At this level of light intensities, most probably,
saturation problem of photo-detectors will be encountered \cite{Lun17,Lun20}.

\begin{figure}
  \centering
  \includegraphics[scale=0.5]{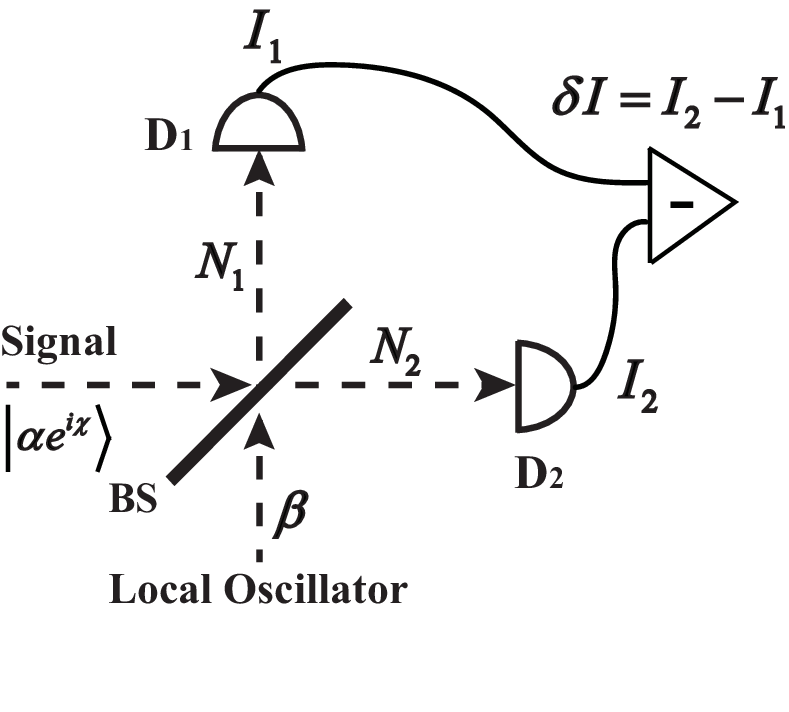}
  \caption{
Schematics for optical phase estimation via homodyne measurement.
The signal light is mixed with a reference light (local oscillator)
through a 50/50 beam splitter.
In linear response regime, field quadrature of the signal light
can be obtained from the difference of currents 
measured in the detectors $D_1$ and $D_2$.      }
\end{figure}

In Refs. \cite{Lun17,Lun20}, the effect of photodetector's saturation
on the precision measurement of deflection of an optical beam
was analyzed in theory and demonstrated by experiment.
In order to avoid the saturation problem,
the strategy of {\it postselection} was proposed,
which guarantees  
the smaller number of postselected photons containing
almost all the metrological information of all the photons
before postselection \cite{AAV88,Jor14,Li20,Li24,Llo20,Ste22,Yang24}.
As a result, the postselection technique provides an intriguing approach
to ensure photodetectors 
operating under the saturation threshold,
even in the case of a large number of probe photons.
In this work, in the absence of postselection,
as in most cases (in the so-called `conventional' measurements),
we generalize the theory 
of extracting the optical phase in homodyne measurement
to the presence of saturation effect of photodetectors.

Let us start with the simple theory of quadrature measurement
of an optical field, as schematically shown in Fig.\ 1.
The phase parameter $\chi$ to be estimated is encoded 
in the single-mode coherent state (signal light) $|\alpha e^{i\chi}\ra$.
For the purpose of homodyne measurement,
a reference light (usually called local oscillator), 
with the same frequency of the signal light 
and a complex amplitude
$\beta= i|\beta| e^{i\varphi}$ (the phase can be modulated), 
is introduced to mix with the signal light
through a $50/50$ beam splitter \cite{WM09}.
Then, the optical fields to be detected by the detectors $D_1$ and $D_2$
are $|\widetilde{\alpha}_1\ra$ and $|\widetilde{\alpha}_2\ra$,
with $\widetilde{\alpha}_1=(\beta + i\alpha)/\sqrt{2}$
and $\widetilde{\alpha}_2=(\alpha + i\beta)/\sqrt{2}$.
The intensities of the two optical pulses
are characterized by the mean photon numbers
$N_1=|\widetilde{\alpha}_1|^2$ and $N_2=|\widetilde{\alpha}_2|^2$.
The electric currents in $D_1$ and $D_2$
are proportional to the incident photon numbers,
which can be denoted as
$I_1=r N_1$ and $I_2=r N_2$,
with $r$ the opto-electric conversion constant.
From the difference of $I_2$ and $I_1$, we can obtain
\bea\label{d-current}
\delta \tilde{I} =\frac{I_2-I_1}{2|\beta| r}
= |\alpha| \sin(\chi-\varphi) \,.
\eea
Actually, this is the result of the quantum average of
the field quadrature operator
$X=-i(a e^{-i\varphi}-a^{\dagger} e^{i\varphi})/2$
in the state $|\alpha e^{i\chi}\ra$.
This result is also obtained from modulating a relative phase 
$\varphi+\pi/2$ of the local oscillator (LO) with respect to the signal light.
One can extract the unknown phase $\chi$ from the above difference current.
This is the basic principle of optical phase estimation via homodyne measurement.

Below, following Refs.\ \cite{Lun17,Lun20}, we present a theory of photodetectors.
In the linear response regime, it can recover the above result of
standard homodyne measurement. However, we will show that
the above result cannot be applied in nonlinear response regime,
which corresponds to the presence of saturation effect.
Specifically, let us consider first $n$ photons being injected into the detector $D_j$.
The responding number of photoelectrons in the detector is a random variable, 
which is assumed to satisfy the Gaussian probability distribution
\bea
P_j(k|n) = (2\pi\sigma_j^2)^{-1/2} e^{-[k-\mu_j(n)]^2/2\sigma_j^2} \,.
\eea
Following Refs.\ \cite{Lun17,Lun20},
the average number of photoelectrons can be modeled as
\bea\label{n-average}
\mu_j(n) = k^{(j)}_{\rm max}\left(1- e^{-n/N^{(j)}_{\rm sat}}\right) \,,
\eea
with $k^{(j)}_{\rm max}$ the opto-electric converting factor,
and $N^{(j)}_{\rm sat}$ the threshold parameter of saturation effect.

Now consider the optical field entering the $j_{\rm th}$ detector,
$|\widetilde{\alpha}_j\ra = \sum_n c_{j,n} |n\ra$.
The photon number in this state is uncertain, 
which satisfies the Poissonian distribution
$P_j(n)=|c_{j,n}|^2=e^{-N_j} N_j^n/n!$, where $N_j=|\widetilde{\alpha}_j|^2$.
Then, the responding electron number in the detector
satisfies the following joint probability distribution
\bea\label{Pjk}
\widetilde{P}_j(k) = \sum_{n=0}^{\infty} P_j(k|n) P_j(n)  \,.
\eea
Based on this probability function, one can evaluate the averages of $k$ and $k^2$.

In the linear response regime, $\mu_j(n)$ can be approximated as
$\mu_j(n)\simeq k^{(j)}_{\rm max}/N^{(j)}_{\rm sat} n \equiv \tilde{r}_j n$.
Then, we have $I_j = e\,\tau_w^{-1}\sum_k k \widetilde{P}_j(k)= r_j N_j$,
with $r_j$ defined by $r_j=e\,\tilde{r}_j/\tau_w$.
Here, $e$ is the electron charge and $\tau_w$ 
the response window time of the detector.
Assuming that the opto-electric response coefficients of $D_1$ and $D_2$
are the same, $r_1=r_2=r$,
the result of \Eq{d-current} is then recovered, i.e., $N_2-N_1=(I_2-I_1)/r$.

In the nonlinear response regime, using $\widetilde{P}_j(k)$,
we can also carry out an analytic expression for the average of $k$,
thus the electric current in the detector, as follows
\bea\label{I-average}
I_j &=& e\, \tau_w^{-1}\sum_k k \widetilde{P}_j(k) = e\, \tau_w^{-1}\sum_n \mu_j(n) P_j(n)  \nl
&=& I^{(j)}_{\rm max} \left( 1- e^{-N_j/\widetilde{N}^{(j)}_{\rm sat} } \right) \,.
\eea
Here we introduced
$I^{(j)}_{\rm max}=e\, k^{(j)}_{\rm max}/\tau_w$
and $\widetilde{N}^{(j)}_{\rm sat} = (1- e^{-1/N^{(j)}_{\rm sat}})^{-1}$.
Since $N^{(j)}_{\rm sat}\gg 1$,
$\widetilde{N}^{(j)}_{\rm sat}$ can be well approximated by $N^{(j)}_{\rm sat}$.
Then, comparing \Eqs{n-average} and (\ref{I-average}),
we have $\la \mu_j(n)\ra = \mu_j(\la n\ra)$.
Here $\la \cdot\ra$ means an average over the distribution function $P_j(n)$.
This relation is not obvious at all {\it in priori}.

Based on \Eq{I-average}, we can introduce an inverse function ${\cal F}$
and reexpress \Eq{I-average} as $N_j={\cal F}(I_j)$. Then we have
\bea\label{IN-diff}
{\cal F}(I_2)-{\cal F}(I_1) = N_2 - N_1 = 2|\alpha \beta| \sin(\chi-\varphi) \,.
\eea
Using this formula, one can extract the unknown phase $\chi$,
based on the measured average currents $I_1$ and $I_2$
in the nonlinear response regime,
which corresponds to the presence of saturation effect.
The saturation effect will become prominent 
when $N_j$ is comparable to or larger than $N^{(j)}_{\rm sat}$.

\begin{table}[H]
  \begin{center}
  \caption{ Comparison between two treatments for the responding photoelectrons
and photocurrent in a photodetector,
which has $N_{\rm sat}=10^{17}$, $k_{\rm max}/N_{\rm sat}=0.1$,
and response time window $\tau_w=0.1$ ms.
The results $\bar{k}^{({\rm L})}$ and $\bar{I}^{({\rm L})}$
are from the linear response approximation treatment,
while $\bar{k}^{({\rm nL})}$ and $\bar{I}^{({\rm nL})}$
from nonlinear response treatment.}
    \label{tab_table}
    \renewcommand\arraystretch{1.5}
    \newcolumntype{L}{>{\centering\arraybackslash}p{1cm}}
    \newcolumntype{C}{>{\centering\arraybackslash}p{1.7cm}}
    \newcolumntype{D}{>{\centering\arraybackslash}p{1.2cm}}
    \newcolumntype{E}{>{\centering\arraybackslash}p{2cm}}
    \newcolumntype{F}{>{\centering\arraybackslash}p{1cm}}
    \begin{tabular}{L C D E F}
    \hline
    \hline
      \multicolumn{1}{l}{\textbf{$N/{N_{\rm sat}}$}} &
      \multicolumn{1}{c }{${{\bar k}^{({\rm L})}}/{k_{\max }}$} &
      \multicolumn{1}{c }{${I^{({\rm L})}}(\mathrm {A})$} &
      \multicolumn{1}{c }{${{\bar k}^{({\rm nL})}}/{k_{\max }}$} &
      \multicolumn{1}{c }{${I^{({\rm nL})}}(\mathrm{A})$} \\
      \hline
      \textbf{$0.01$}  & $0.01$    & $0.16$    & $0.01 $   &  $0.16$ \\

        \textbf{$0.1$} & $0.1$   & $1.60$    & $0.1$   & $1.53$  \\

      \textbf{$1$}  & $1$   & $16.02$   & $0.63$    & $10.13$  \\

      \textbf{$2$}  & $2$   & $32.04$   & $0.87$    & $13.85$  \\

      \textbf{$3$}  & $3$   & $48.07$   & $0.95$    &  $15.22$  \\
      \hline
      \hline
    \end{tabular}
  \end{center}
\end{table}

\begin{figure}
  \centering
  \includegraphics[scale=0.45]{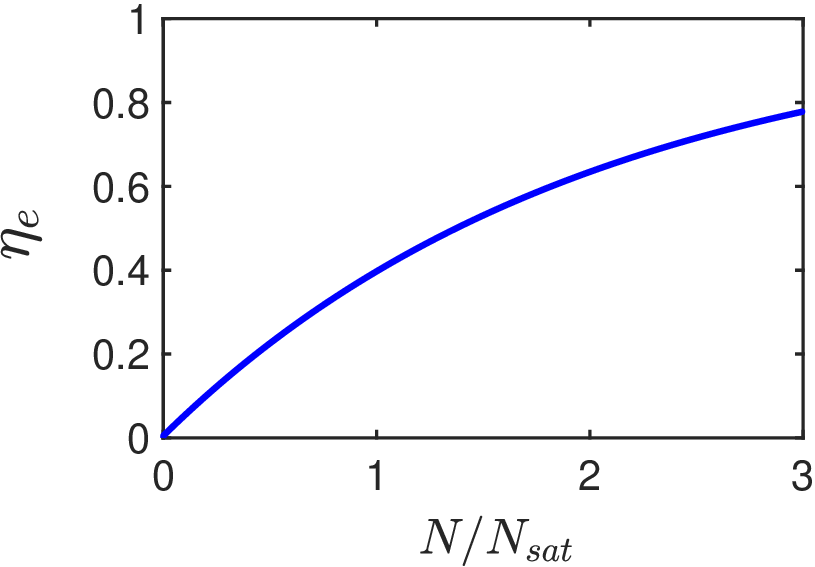}
  \caption{
Error ratio $\eta_e=|\widetilde{\chi}-\chi |/\chi$
{\it versus} the average photon number $N$ of the signal light.
$\eta_e$ characterizes the deviation
of the estimated phase $\widetilde{\chi}$ from the real $\chi$.
Parameters used:
$\chi=0.01$, $N_{\rm sat}=10^{17}$, $k_{\rm max}/N_{\rm sat}=0.1$,
and $|\beta|^2=10^{15}$.    }
\end{figure}

Below, using specific parameters
consistent with the estimation in Refs.\ \cite{Ra17,Sab17,Sab18}, 
we illustrate the effect of saturation
and the improved estimation of $\chi$ in the nonlinear response regime,
by applying the above results of \Eqs{I-average} and (\ref{IN-diff}).
Let us consider a continuous wave (CW) laser with power $P=20$ W
and at central frequency $\omega=100$ THz,
and assume the response time window of a photodetector $\tau_w=0.1$ ms.
One can estimate the photon number $N=P\tau_w/\hbar\omega\simeq 10^{16}$
for each time of detection by the detector.
We assume the saturation threshold parameter $N_{\rm sat}=10^{17}$
and the opto-electric response coefficient $k_{\rm max}/N_{\rm sat}=0.1$.
The average electron number,
$\bar{k}=\sum_k k \widetilde{P}_j(k)$,
is roughly $\bar{k}\simeq 10^{15}$.
Then, we may estimate the electric current in the detector as
$I=e\, \bar{k}/\tau_w \simeq 1.6$ A.
This estimation is for the linear response regime,
in the absence of saturation effect.
In Table I, we display a few more results to show how a saturated current
will be reached, with the increase of light intensity,
after passing through a nonlinear response regime.
The results in Table I also indicate that, in the nonlinear response regime,
the saturation effect will seriously affect the estimation, 
if we still apply the usual linear response protocol, i.e., based on \Eq{d-current}.

For instance, let us assume $N_{\beta}=|\beta|^2=10^{15}$ photons
(during the time interval $\tau_w=0.1$ ms) for the intensity of the LO.
In Fig.\ 2 we show the error caused by increasing $N$
over the threshold $N_{\rm sat}$,
if we apply the standard protocol, \Eq{d-current}. 
In this plot, for the estimation of the optical phase $\chi$,
we have defined the error ratio,
$\eta_e=| \widetilde{\chi}-\chi |/\chi$,
which characterizes the deviation
of the estimated phase $\widetilde{\chi}$ from the real $\chi$.
We may point out that, roughly speaking, the nonlinear response regime
is restricted by $N$ smaller than a few $N_{\rm sat}$.
Beyond this regime, the detectors do not respond to more photons,
with the currents being saturated at $I_{\rm max} = e k_{\rm max}/\tau_w$.
In this case, based on the measured currents $I_1$ and $I_2$ ,
even using \Eqs{I-average} and (\ref{IN-diff}),
one cannot extract the phase $\chi$.

The above estimation is based on the average currents $I_1$ and $I_2$,
from quantum ensemble measurements.
For each single shot measurement, using a single $|\alpha e^{i\chi}\ra$,
the output currents $i_1$ and $i_2$ are random variables, 
having deviations $\delta I_1=i_1-I_1$ and $\delta I_2=i_2-I_2$ from the average currents.
One may characterize in theory the estimate precision of $\chi$ owing to such uncertainties.
Let us denote ${\cal F}(I_2)-{\cal F}(I_1)={\cal M}(I_1,I_2)$.
Then, from the rule of error propagation, we have
$\delta {\cal M} = [(\frac{\partial{\cal M}}{\partial I_1})^2 \delta^2I_1
+ (\frac{\partial{\cal M}}{\partial I_2})^2 \delta^2I_2  ]^{1/2}$.
Based on \Eq{IN-diff}, we further obtain
\bea
&& ~~~~ |\frac{\partial(N_2-N_1)}{\partial \chi}|\, \delta\chi
= 2|\alpha\beta\cos(\chi-\varphi)|\, \delta\chi   \nl
&&= \left[\left(\frac{d{\cal F}(I_1)}{d I_1}\right)^2 \delta^2I_1
+ \left(\frac{d {\cal F}(I_2)}{d I_2}\right)^2 \delta^2I_2  \right]^{1/2}
\eea
Using $\widetilde{P}_j(k)$, given by \Eq{Pjk}, we obtain the variance
$\delta^2 k_j = \sum_k k^2 \widetilde{P}_j(k) - \bar{k}_j^2  = \sigma_j^2$,
and thus $\delta^2 I_j =(\frac{e}{\tau_w})^2 \sigma_j^2$.
Using \Eq{I-average}, we also obtain
$d{\cal F}(I_j)/d I_j= \widetilde{N}^{(j)}_{\rm sat} \, (I^{(j)}_{\rm max} - I_j)^{-1}$.
For simplicity, we consider two identical detectors, by assuming
$I^{(1)}_{\rm max}=I^{(2)}_{\rm max}\equiv I_{\rm max}$,
$\sigma_1=\sigma_2\equiv \sigma$,
and $\widetilde{N}^{(1)}_{\rm sat} = \widetilde{N}^{(2)}_{\rm sat}
\equiv \widetilde{N}_{\rm sat}$. We finally obtain
\bea\label{precision}
\delta\chi = \frac{e \sigma \widetilde{N}_{\rm sat}}
{2\tau_w\, |\alpha \beta \cos(\chi-\varphi)|}
\left(\Delta_1^2 + \Delta_2^2\right)^{1/2}
\eea
Here we introduced $\Delta_j = (I_{\rm max}-I_j)^{-1}$.
Notice that $|\alpha|=\sqrt{N}$.
This result implies that the estimate precision
scales with the average photon number
by $\sim 1/\sqrt{N}$. This is the SQL, as expected.
Moreover, in practice,
the estimation is based on the results of multiple measurements,
e.g., $M$ times repetitive measurements.
Thus, the variance of the responding electron number in the detector
will be reduced from $\delta^2 k_j =\sigma^2$ to $\sigma^2/M$.
Therefore, the estimate precision $\delta\chi$ given by \Eq{precision}
scales with the photon numbers and measurement times as $\sim 1/\sqrt{MN}$.
In Refs.\ \cite{Ra17,Sab17,Sab18}, measurement times $M=10^9$ and $10^{10}$
are considered.

To summarize, we have presented a theory which generalizes
the optical phase estimation via homodyne measurement
from detector's linear to nonlinear response regime.
For optical coherent light, we carried out analytic expressions
for detector's current and estimate precision of the optical phase.
The generalized theory can work in the presence of saturation effect.
However, it should be noted that in the oversaturation regime,
the photodetector does not respond to more photons.
In this case, one is unable to extract the phase parameter,
since the currents do not carry information of the phase.
In order to avoid the oversaturation problem,
the technique of postselection should be a promising strategy,
as proposed and demonstrated in Refs.\ \cite{Lun17,Lun20}.
For the possible detection of relativistic gravity effects
studied in Refs.\ \cite{Ra17,Sab17,Sab18},
based on nonlinear laser interferometer and field quadrature measurement,
most probably, the required high-intensity lasers
will cause oversaturation problem of photodetectors.
As analyzed in our recent work \cite{Li24},
applying the postselection technique,
one can ensure the detector being operated
under the saturation threshold, even using high-intensity lasers.
The most remarkable advantage of postselection technique
is that, the smaller number of postselected photons can contain
almost all the metrological information of all the photons before postselection.

\vspace{0.5cm}
{\flushleft\it Acknowledgements.}---
This work was supported by the NNSF of China (Nos.\ 11675016, 11974011 \& 61905174).


\end{CJK}
\end{document}